\documentclass{article}
\usepackage{graphicx} 
\usepackage[letterpaper, margin=1in]{geometry}
\usepackage[numbers]{natbib}
\usepackage{float}
\usepackage{lscape}
\usepackage{wrapfig}
\usepackage{authblk}
\usepackage{tabularx}
\usepackage{aas_macros}

\title{ {\Huge Ardua}  \\ Unveiling the Baryon Cycle from Stars to the Cosmic Web \\ {\large \textit{(NASA ASTRA Initiative concept submitted to COPAG and PhysPAG)}}}

\author[1]{Carlos J. Vargas}
\author[2]{Caroline Kilbourne}
\author[1]{Haeun Chung}
\author[1]{Erika Hamden}
\author[3]{Ralph Kraft}
\author[4]{Joseph N. Burchett}
\author[1]{Lauren Corlies}
\author[5]{Claude-Andr\'e Faucher-Gigu\`ere}
\author[6]{Kevin France}
\author[7]{Keri Hoadley}
\author[6]{Briana Indahl}
\author[3] {Dong-Woo Kim}
\author[8]{Varsha Kulkarni}
\author[9]{Jiangtao Li}
\author[10]{Nicole Melso}
\author[11]{Drew Miles}
\author[12]{Nikole M. Nielsen}
\author[2,20]{Anna Ogorzalek}
\author[13]{Ben Oppenheimer}
\author[14]{Frits Paerels}
\author[3]{Daniel Patnaude}
\author[15]{Molly Peeples}
\author[2]{Frederick S. Porter}
\author[14]{David Schiminovich}
\author[3] {Malgorzata Sobolewska}
\author[16]{Ming Sun}
\author[17]{Todd Tripp}
\author[15]{Jason Tumlinson}
\author[18]{Sarah Tuttle}
\author[18]{Jessica Werk}
\author[19] {Ka-Wah Wong}
\author[3] {John ZuHone}

\affil[1]{Department of Astronomy \& Steward Observatory, University of Arizona, Tucson, AZ, USA}
\affil[2]{NASA Goddard Space Flight Center, Greenbelt, MD, USA}
\affil[3]{Center for Astrophysics, Harvard \& Smithsonian, Cambridge, MA, USA}
\affil[4]{Department of Astronomy, New Mexico State University, Las Cruces, NM, USA}
\affil[5]{CIERA and Department of Physics and Astronomy, Northwestern University, Evanston, IL, USA}
\affil[6]{Laboratory for Atmospheric and Space Physics, University of Colorado Boulder, Boulder, CO, USA}
\affil[7]{Department of Astronomy, University of Florida, Gainesville, FL, USA}
\affil[8]{Department of Physics and Astronomy, University of South Carolina, Columbia, SC, USA}
\affil[9]{Purple Mountain Observatory, Chinese Academy of Sciences, Nanjing, China}
\affil[10]{School of Physics and Astronomy, Rochester Institute of Technology, Rochester, NY, USA}
\affil[11]{California Institute of Technology, Pasadena, CA, USA}
\affil[12]{Homer L. Dodge Department of Physics and Astronomy, University of Oklahoma, Norman, OK, USA}
\affil[13]{Center for Astrophysics and Space Astronomy, University of Colorado Boulder, Boulder, CO, USA}
\affil[14]{Department of Astronomy, Columbia University, New York, NY, USA}
\affil[15]{Space Telescope Science Institute, Baltimore, MD, USA}
\affil[16]{Department of Physics and Astronomy, University of Alabama Huntsville, Huntsville, AL}
\affil[17]{Department of Astronomy, University of Massachusetts Amherst, Amherst, MA, USA}
\affil[18]{Department of Astronomy, University of Washington, Seattle, WA, USA}
\affil[19]{Department of Physics, The State University of New York – Brockport, Brockport, NY, USA}
\affil[20] {Center for Research and Exploration in Space Science and Technology, NASA / GSFC (CRESST II), Greenbelt, MD, USA}

\begin{document}

\maketitle

\newpage
\vspace{-3mm}
\section{Science Investigation}

\vspace{-2.5mm}
\subsection{The Baryon Cycle: The Missing Piece of Galaxy Evolution}
\vspace{-1.5mm}

Every galaxy in the universe is governed by a complex cycle, where the very fuel for its sustenance is mixed with cosmic debris from the formation of stars. Gas falls in from the cosmic web, condenses into stars, and is returned to the surrounding medium through stellar winds, supernovae, and feedback from accreting supermassive black holes. This baryon cycle — the continuous exchange of gas, metals, and energy between galaxies and their halos — is the engine of cosmic evolution. It determines how fast galaxies grow, when they stop forming stars, and what they look like today. Yet despite decades of progress, the baryon cycle remains the most poorly constrained facet of galaxy formation. The reason is simple: most of the cycle happens in the circumgalactic medium (CGM), and the CGM has never been systematically mapped.

The CGM is the vast reservoir of diffuse, multiphase gas extending hundreds of kiloparsecs around every galaxy. The CGM serves as the intermediary between galaxies 
and the intergalactic medium. Absorption-line surveys with HST/COS have established that this reservoir is massive, containing comparable mass to the 
stars in warm ionized gas alone, and that it holds the majority of a galaxy's metals (e.g., \citep{tumlinson17}). 
The low density of the CGM makes it particularly difficult to detect in emission. But the CGM is not a single uniform medium. It spans several orders of magnitude in temperature simultaneously, from shielded cold gas filaments at several hundred K, to warm ionized gas at $10^4$ K, to hot X-ray emitting plasma at $10^6$ -- $10^7$ K, with each phase playing a distinct role in the baryon cycle and each phase requiring different observational tools to detect. The cold and warm phases, accessible in the ultraviolet, may carry much of the CGM's energy and angular momentum and trace the flows of gas into and out of galaxies. The hot phase, accessible only in the X-ray, may dominate the CGM mass in halos around the Milky Way mass scale and above, and may interact with and confine cool clouds. 
No phase can be understood in isolation. The baryon cycle involves all of these phases and their interactions, and understanding its complexities requires \textit{comprehensive} access to the full range of CGM gas phases, which has not yet been possible with existing UV and X-ray missions. \textbf{This is the central scientific problem Ardua is designed to solve.}

\vspace{-4mm}
\subsection{UV and X-Ray: Each Incomplete Without the Other}\label{sec:UVandXray}
\vspace{-2mm}

The UV bandpass contains spectral emission lines tracing every phase of the CGM below T $\sim10^6$ K. Ly$\alpha$ (1216 \AA) provides the brightest available tracer of cool gas spatial extent. C II (1335 \AA), and Si II (1260 \AA) trace the cool ionized phase at T $\sim10^4$ -- $10^5$ K. Fluorescence lines including many H$_2$ features at $912$ -- $1700$ \AA, and C II*, Si II* features in the $1200$ -- $1600$ \AA~range provide additional cool phase diagnostics and have never been mapped systematically in nearby galaxies. C IV (1550 \AA) and Si IV (1394 \AA) trace the warm phase at T $\sim10^5$ K. Higher ions such as O VI (1032/1038 \AA) and Ne VIII (770/780 \AA) trace the warm-hot phase at T $\sim3\times10^5$ -- $10^6$ K — this phase can either be volume-filling around Milky Way-mass galaxies or trace the 
turbulent mixing layers between cool and hot gas, a long-standing question that emission mapping is required to answer.
This list spans over three decades in temperature, all accessible utilizing existing technology advancements in UV optics and detectors.

The X-ray bandpass picks up where the UV leaves off. O VII at 0.57 keV, O VIII at 0.65 keV, and Fe XVII at 0.73 keV trace coronal gas at T $\sim10^6$ -- $10^7$ K — the volume-filling hot phase that dominates the pressure support budget of virialized halos. A single spatially resolved detection of O VII emission from a MW-mass galaxy halo would be transformative, providing the first direct observational confirmation that a volume-filling hot corona exists. Combined with simultaneous UV maps of the cool and warm phases in the same halo, the X-ray detection would enable a direct test of thermal pressure balance between phases and the first complete census of baryons across all CGM phases in individual galaxy systems.

Does feedback heat the CGM uniformly or drive multiphase winds? Is the CGM in thermal pressure balance across phases? How does energy deposited by supernovae and AGN partition between heating the hot phase and accelerating cool outflows? The most pressing scientific questions in this field are fundamentally unanswerable from either wavelength range alone. \textbf{Ardua addresses these questions by comprehensively mapping all CGM phases for the first time, using wide-field, high sensitivity X-ray and UV instruments designed for efficient mapping of a large sample of galaxies.} 

\vspace{-4mm}
\subsection{The Key to Distinguishing Between Galaxy Formation Theories\label{sec:sims}} 
\vspace{-2mm}

Cosmological simulations and semi-analytic models of galaxy formation 
have converged on 
galaxy properties that broadly match observations (\citep{somerville15,crain23}). But these same models make dramatically different, often contradictory predictions for the CGM. 
Essentially all the models (except for variations in which non-thermal pressure from cosmic rays is important) predict that the volume-filling phase is hot around galaxies of mass similar to the Milky Way and above, while cooler gas likely dominates at lower masses (e.g., \citep{birnboim03}). 
Interestingly, the different models make different predictions for observations such as X-ray surface brightness profiles, morphologies, and kinematics, which are sensitive to key uncertain processes such as stellar and AGN feedback (e.g., \citep{silich25}). Despite their differences, most simulations predict CGM emission in the UV and X-ray at levels detectable by Ardua. 
Furthermore, the hot and cold CGM play different roles in essential aspects of galaxy formation, such as the formation of disk galaxies and eventually the quenching of star formation, in different models \citep{faucher-giguere23}. 
Unfortunately, existing empirical constraints on the crucial hot phase are presently extremely limited due to the sensitivity of existing X-ray telescopes. 
Observations of the hot gas around typical galaxies
is a critical missing piece to distinguish between galaxy formation and CGM theories. 

Reconciling differences in theoretical prescriptions of galaxy formation and evolution requires direct observational constraints on CGM morphology. To date, the vast majority of observational constraints have come from absorption line spectroscopy. These studies pioneered our knowledge of the CGM but are geometrically limited to ``pencil beam" column density measurements in sightlines that happen to have background light sources. Advancement in galaxy evolution requires a shift from sightline-limited absorption spectroscopy to volume-filled emission mapping of each CGM phase in both the ultraviolet and X-ray.

\vspace{-4mm}
\subsection{Why Now, and Why ASTRA?}
\vspace{-2mm}

The baryon cycle sits at the center of the Astro2020 Decadal Survey's Pathways to Discovery in Astronomy and Astrophysics for the 2020s \cite{astro2020}, which identifies it explicitly within the priority area ``Unveiling the Drivers of Galaxy Growth." This priority is reinforced by the 2014 NASA Astrophysics Roadmap, Enduring Quests, Daring Visions, which calls for understanding how galaxies acquire and expel gas across cosmic time, and by multiple NASA mission concept studies — including LEM and HabEx — that identified wide-field UV and X-ray spectroscopy of the CGM and intergalactic medium (IGM) as defining science drivers for the next generation of space observatories. No approved or planned mission is designed to obtain deep, comprehensive UV and X-ray emission maps of the same galaxies at survey scale. The Habitable Worlds Observatory (HWO), by contrast, is built for a different problem: its coronagraphic design drives a narrow field of view — a notional $\sim$3$\times$3 arcmin$^2$ MOS FoV — that would require more than 10 deep tiled pointings to map a single halo, incompatible with surveying a large galaxy sample.


Ardua opens an observational landscape the ASTRA Initiative is designed to enable. ASTRA calls for large strategic mission concepts that pursue future portfolio balance, address high-priority science identified by the community, and go beyond what existing or approved missions can deliver. Ardua answers all three criteria directly, addressing the highest-priority open question in galaxy formation that no approved mission can fully solve, with an architecture specifically enabled by the maturation of UV detector/coating technology, X-ray microcalorimeter heritage, and sensitivity requirements that are both simulation-derived and soon to be flight-verified by the Aspera Pioneers Mission. Without Ardua, the transformative gas cycling science called for in Astro2020 will remain incomplete throughout HWO's operational lifetime.


\vspace{-4mm}
\subsection{The Baryon Cycle Across Cosmic Scales}
\vspace{-2mm}

The CGM survey program described above represents a primary science driver, but the same instrument capabilities -- wide-field UV and X-ray spectroscopy at unprecedented surface brightness sensitivity -- open a broad range of groundbreaking science. On intergalactic scales, Ardua's UV coverage of O VI and C IV combined with X-ray mapping of OVII and O VIII traces the intergalactic medium across the full temperature range of the WHIM, directly measuring how galaxies have enriched the cosmic web with metals. At cluster outskirts, where infalling WHIM meet the virial shock, O VII and O VIII lines in non-equilibrium ionization (NEI) conditions capture the thermalization of accreting gas in real time, providing a direct observational window into the physics of large-scale structure assembly \cite{Yoshikawa06,Wong11}. On galaxy scales, Ardua will probe UV and soft X-ray emission around the central and satellite galaxies in galaxy groups/clusters to study radiative cooling, turbulent mixing, and AGN feedback. On smaller scales, Ardua maps how stellar feedback shapes the multiphase structure of the ISM and disk-halo interface, tracing the energy and mass exchange between hot superbubble interiors and their cool surroundings. Finally, Ardua's UV and X-ray capability addresses a critical preparatory science gap for HWO: characterizing the FUV and X-ray radiation environments of exoplanet host star candidates, including accretion and outflow physics in young stellar systems, on timescales from minutes to hours that require multi-wavelength observations no other approved mission can provide.

\newpage

\section{Science Traceability}
\vspace{-4mm}
\begin{figure}[H]
 \hspace{-1mm} \includegraphics[width=165mm]{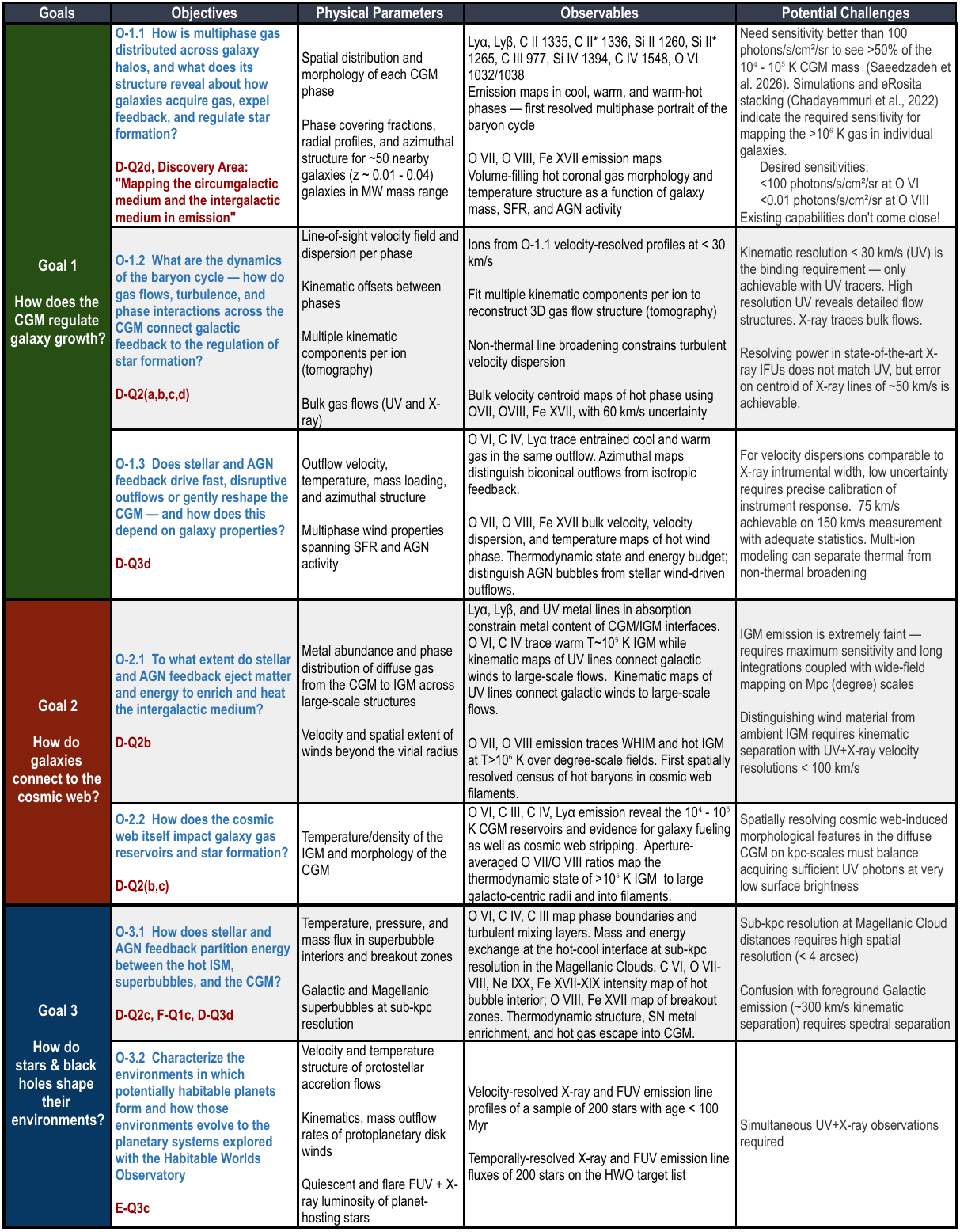}  \caption{\small Science Table. Cited references:\cite{Saeedzadeh_2026}\cite{Chadayammuri_2022}. Codes in the ``Objectives" column refer to Astro2020 Panel Report Discovery/Science Questions (panel letter, question number, sub-letter, e.g., D-Q2a = Panel on Galaxies, Question 2a).}
\end{figure}
\vspace{-4mm}
\section{Instrument Description}
\vspace{-3mm}
\subsection{UV Instrument}
\vspace{-2mm}

Ardua's notional UV instrument concept multiplexes fast, far-UV telescopes feeding compact spectrographs, exploiting a key property of diffuse-source survey speed: for extended low-surface-brightness emission like the CGM, survey speed scales with the detector area and the inverse of the focal ratio, not the aperture size. Therefore, an array of N small, fast telescopes can outperform the survey speed of one large telescope.
A large ($>3$ m), slow (F/\#
$>$16) single-mirror observatory would require over 10
 years to map the CGM of even a single galaxy halo ($>$ 15' $\times$ 15') at the required sensitivity, demonstrating that fast optics are essential regardless of how that \'etendue is partitioned. 


\begin{wraptable}{R}{0.5\textwidth}
\centering
\vspace{-0.5cm}
\includegraphics[width=0.5\textwidth]
{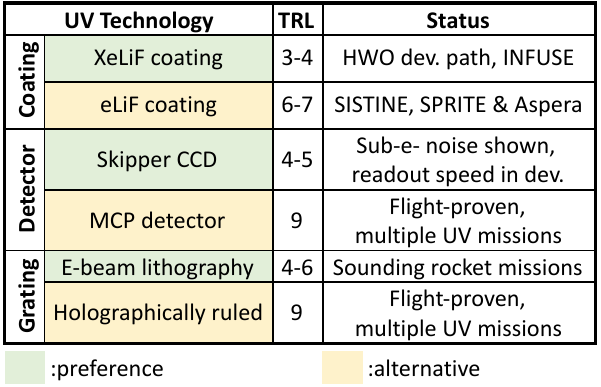}
\vspace{-0.9cm}
\caption{
UV instrument technology readiness, baselining XeLiF coatings, Skipper CCDs, and E-beam lithography gratings.}
\label{tab:uv}
\end{wraptable}


What remains a key open trade is how to partition that \'etendue: a single fast telescope with a large-optics/detector format spectrograph or IFU, an intermediate number of medium telescopes, or many small telescopes each with a compact spectrograph, following the precedent of multiplexed arrays like the Dragonfly Telephoto Array \cite{abraham14}. Back-end spectrograph optical design, fabrication and qualification are likely the dominant cost driver. The study phase will explore this trade space and determine the most cost-effective architecture, with multiple credible options capable, in principle, of reaching the required sensitivity of $10^{-20}$ erg/s/cm$^2$/arcsec$^2$ at O VI, with R$\sim$50,000 over a $>$ 15' $\times$ 15' FOV. The wavelength range would cover various FUV emission lines (Ly$\alpha$, Ly$\beta$, O VI, C II, C III, C IV, Si IV) from nearby galaxy halo.

\vspace{-0.4cm}
\paragraph{Technology readiness and maturation.} No element requires invention. Primary maturation items: (1) Skipper CCD readout speed for survey-mode cadence, (2) scale-up and space qualification of XeLiF/AlF$_3$ coatings, and (3) system-level control of FUV stray light and diffuse foreground backgrounds, which set the sensitivity floor independent of detector performance.

\vspace{-4mm}
\subsection{X-ray Instrument}
\vspace{-2mm}

The X-ray instrument is based on the Line Emission Mapper (LEM) X-ray probe concept \cite{kraft24}. A large-area grazing-incidence optic with 4-meter focal length focuses onto a $>$30' wide array of 15" TES microcalorimeter pixels optimized for the 0.2--2 keV band; a $>$5
'$\times$5' inner array achieves $<$1.3 eV FWHM resolution, with the surrounding outer array trading resolution for focal-plane coverage at $<$2.5 eV FWHM. This short, fast telescope configuration maximizes field of view and spectral resolution for CGM science and represents a different optimization from New Athena's X-IFU. The LEM requirements serve as the foundation for Ardua's X-ray instrument, from which targeted trade studies will be performed.

\begin{wraptable}{R}{0.5\textwidth}
\centering
\vspace{-0.5cm}
\includegraphics[width=0.5\textwidth]
{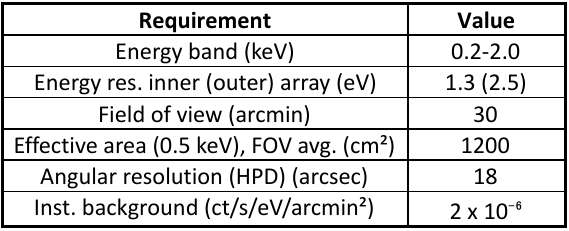}
\vspace{-0.9cm}
\caption{
The LEM requirements, which form the foundation for the Ardua X-ray instrument}
\label{tab:xray}
\end{wraptable}


\vspace{-0.4cm}
\paragraph{Technology readiness and maturation needs.} LEM technologies were assessed at TRL 5 at the time of the Step-1 Probe submission. The baseline grazing-incidence hyperboloid-hyperboloid optic uses coated single-crystal silicon reflectors; angular resolution requirements for a CGM mission are modest, driven by point-source removal rather than arcsecond imaging. The spectrometer follows X-IFU baselines closely, providing high confidence in the reference capabilities. The primary maturation need is demonstration of the optic at relevant scale.

\clearpage
\section{Mission Implementation}
\vspace{-3mm}
The Ardua mission architecture will be designed to support the high level observing program, as well as be flexible enough to provide a community resource. The architecture described here is notional, but will be refined through a study phase to ensure scientific success and low technical risk.

Ardua's observing program is focused on nearby galaxy targets, exoplanet hosts, the Magellanic stream, and more. As a baseline mission, we plan to survey $>$50 nearby galaxies, at distances from 2 to 100 Mpc for CGM mapping, addressing objectives associated with Goal 1. We will conduct a study of $>$100 more distant galaxies (but within z$<$1) which are currently unobserved by visible wavelength surveys for larger spatial scale emission mapping of the cosmic web (Goal 2). We will observe the Magellanic Clouds and associated stream, addressing O-3.1. We plan to observe $>$ 200 exoplanet host stars in the HWO target list, to complete O-3.2. Finally, the mission will provide time for community science targets, which will drive the requirements on flexibility of the spacecraft bus (i.e. things like slew rates, frequency of communications, ToO policy, etc). There are no unusual data processing requirements.

The UV and X-ray sky backgrounds are intrinsically very low. The UV background in low earth orbit (LEO) is driven primarily by the atmosphere (Oxygen and Hydrogen lines in the FUV). Because of this, our preference is either L2 or a high Earth Orbit (HEO), similar to TESS or the orbit planned for UVEX. The HEO orbit gives a balance between low background levels, good bandwidth for up/downlink, and moderate cosmic ray rates. In either case, radiation tolerance in these higher orbits will drive requirements on detector shielding and other component level decisions.  For the X-ray instrument, the quiescent unrejectable instrument background that results from cosmic-rays and their secondaries is much lower in low-inclination LEO than outside of the Earth's protective field, but this benefit is countered by reduced observing efficiency and a more challenging thermal environment.  Orbit at L1 was chosen for LEM and is the orbit currently planned for New Athena. L2 was also determined to also be acceptable for LEM. Thus L1 and L2 will be included in the orbit trade study for Ardua.

We conservatively assume a 5 year mission lifetime. This will provide sufficient time to close on our science requirements as well as providing more than two years to community science investigations. The design lifetime of typical spacecraft has been nearly 10 years for most recent missions, and we project we will not require anything particularly unusual.  The cooling chain needed to provide the 40-mK heat sink for the X-ray microcalorimeter will be entirely cryogen-free. The outer shell of the dewar will be $\sim$260 K in orbit and the coolers will operate with it at room temperature, allowing ground testing outside of a thermal-vacuum chamber.


As part of the Ardua study, we will work to understand the pros and cons of 2 separate spacecraft platforms, one for the UV and and one for X-ray instrument. Two platforms provides many advantages- including the ability to conduct science more efficiently (which is valuable for a guest investigator program). But two platforms may add cost. We will investigate cost-reduction approaches such as the use of common bus components.

We plan to leverage the recent rapid development in smaller spacecraft providers to reduce costs. We will also use university-based tests and programs which reduce cost (for example, the University of Arizona can conduct instrument and full integrated spacecraft environmental testing at a fraction of the price of a prime), in addition to having a proven student/university-led software development program for mission operations and data pipeline.  

During the study period, we will explore collaborations with several international partners. Talks will begin with scientific collaborators, and grow from there. Collaboration could include scientific contributions, but also components such as optics, gratings, cryocoolers, etc.

\bibliography{ref}
\bibliographystyle{plainnat}

\end{document}